# SUPERSYMMETRY AND QUANTUM COSMOLOGY

*F. Assaoui* \* and *T. Lhallabi* \*

The Abdus Salam International Centre for Theoretical Physics, Trieste, Italy.

## Abstract

The development of the N = 4 supersymmetric approach to quantum cosmology based on the non-compact global O(d,d) symmetries of the effective action is given. The N = 4 supersymmetric action whose bosonic sector is invariant under O(d,d) is determined. A representation for supercharges is obtained and the form of the zero and one-fermion quantum states leading to the Wheeler–DeWitt equation is found.

---

\* Permanent address: Section of High Energy Physics, H. E. P. L, University MohammedV, Scientific Fac, Rabat, Morocco.
E-mail: Lhallabi@fsr.ac.ma;
E-mail: assfa@usa.net

# 1 - Introduction

The development of the quantum cosmology has been an important motivation in the sense that the initial conditions for the emergence of the Universe as a classical result have been explained. In principle the form of the wave function satisfying the Wheeler-DeWitt equation must be obtained [1]. This equation describes the annihilation of the wave function by the Hamiltonian operator and admits an infinite number of solutions. The boundary conditions must be taken in order to specify the wave function uniquely and are viewed as an additional physical law [2]. On the other hand, in string theory there were many attempts to develop a consistent theory of string cosmology where inflation plays an important role [3]. The solutions of the non-linear sigma model equations in a Friedman-Robertson-Walker background for the graviton, dilaton and antisymmetric tensor have been found [4, 3]. Furthermore, for isotropic Friedman-Robertson-Walker cosmologies, which are restricted to flat space, the dilaton-graviton sector of the string effective action is invariant under an inversion of the scale factor and a shift in the dilaton field [5]. A supersymmetric extension of the quantum cosmology is obtained from the scale factor duality, which is a subgroup of T-duality [6, 7]. Moreover, $N = 2$ supersymmetric approach to quantum cosmology, with spatially flat and homogeneous Bianchi type I Universe admitting d-compact abelian isometries, is developed by employing the non-compact global symmetries of the string effective action [8].

The purpose of the present paper is to develop an $N = 4$ supersymmetric approach to quantum cosmology whose bosonic sector is invariant under global O(d,d) transformations. The outline of this paper is as follows: In section 2, we recall the $N = 2$ supersymmetric approach to quantum cosmology given by J. E. Lidsey and J. Maharana [8] which leads to the O(d,d) invariant Wheeler-DeWitt equation. In section 3, we present the extension of this model to the $N = 4$ supersymmetric case where the bosonic sector is invariant under the global O(d,d) transformations. The corresponding super-constraints on the wave function are then obtained. Thereafter, we derive the classical Hamiltonian by using the classical momenta conjugate to the bosonic and fermionic degrees of freedom. Furthermore, the $N = 4$ supersymmetric quantum constraints are solved for the zero fermion and one-fermion states leading to the Wheeler-DeWitt equation. Finally, in section 4, we make concluding remarks and discuss our results.

## 2 - N = 2 Supersymmetric String Quantum Cosmology

In this section we recall the N = 2 supersymmetric approach to quantum cosmology given by J. E. Lidsey and J. Maharana in ref. [8] where the non-compact global symmetries of the string effective action are used. However, the tree-level string effective action is given by [9].

$$S = \frac{1}{2\lambda_S^{d-1}} \int d^{d+1}\sqrt{|g|}\, e^{-\phi}\left[R + \partial_\mu\phi\, \partial^\mu\phi - \frac{1}{12}H_{\mu\nu\alpha}H^{\mu\nu\alpha} + V\right] \quad (2.1)$$

where $\Phi$ is the dilaton field, $H_{\mu\nu\alpha}$ is the field strength of the antisymmetric torsion tensor $B_{\mu\nu}$, R is the Ricci curvature scalar of the space-time with metric $G_{\mu\nu}$, $g \equiv \det G$, $\lambda_S$ is the fundamental string length scale and V is an interaction potential. The integration over the spatial variables in equation (2.1), by specifying $\lambda_S \equiv 2$ [10], implies that

$$S = \int d\tau \left[\bar{\phi}'^2 + Tr(M'(M^{-1})') + V e^{-2\bar{\phi}}\right] \quad (2.2)$$

where a spatially closed, flat, homogeneous ( Bianchi type I ) space-time is assumed and where the dilaton and two-form potential are taken to be constant on the surfaces of homogeneity
t = constant. Furthermore, the shifted dilaton field $\bar{\phi}$ and the dilaton time parameter are respectively given by

$$\bar{\phi} \equiv \phi - \frac{1}{2}Ln|g| \quad (2.3)$$

$$\tau = \int^t dt_1\, e^{\bar{\phi}(t_1)} \quad (2.4)$$

M is a symmetric 2d × 2d matrix taken as follows:

$$M = \begin{pmatrix} G^{-1} & -G^{-1}B \\ BG^{-1} & G - BG^{-1}B \end{pmatrix} \quad (2.5)$$

where prime denotes differentiation with respect to $\tau$. The matrix M is an element of the group O(d,d) satisfying the conditions

$$M \eta M = \eta \tag{2.6}$$
$$M = M^T$$

and its inverse is given linearly by

$$M^{-1} = \eta M \eta \tag{2.7}$$

where

$$\eta = \begin{pmatrix} 0 & I \\ I & 0 \end{pmatrix} \tag{2.8}$$

and I is the d × d unit matrix. Therefore, the kinetic part of the action (2.2) is invariant under global O(d,d) transformations [10]:

$$\tilde{\bar{\phi}} = \bar{\phi}$$
$$\tilde{M} = \Omega^T M \Omega \tag{2.9}$$
$$\Omega^T \eta \Omega = \eta$$

with $\Omega$ a constant matrix.

On the other hand, the classical Hamiltonian for this cosmological model is given by

$$H_{Bos} = \frac{1}{4} \pi_{\bar{\phi}}^2 - 2 Tr(M \pi_M M \pi_M) - V e^{-2\bar{\phi}} \tag{2.10}$$

where

$$\pi_{\bar{\phi}} = 2 \bar{\phi}'$$
$$\pi_M = -\frac{1}{4} M^{-1} M' M^{-1} \tag{2.11}$$

are respectively the canonical momenta of $\bar{\phi}$ and M. The cosmology is quantified by identifying the momenta (2.11) with the following differential operators:

$$\pi_{\bar{\phi}} = -i \frac{\delta}{\delta \bar{\phi}}, \qquad \pi_M = -i \frac{\delta}{\delta M} \tag{2.12}$$

The insertion of equations (2.12) into the expression (2.10) leads to the Wheeler-DeWitt equation [11].

$$\left[\frac{\delta^2}{\delta\bar{\phi}^2} + 8Tr\left(\eta\frac{\delta}{\delta M}\eta\frac{\delta}{\delta M}\right) + 4V e^{-2\bar{\phi}}\right]\psi(\bar{\phi},M) = 0 \quad (2.13)$$

which is manifestly free from problems of quantum ordering [12]. In order to obtain an N = 2 supersymmetric lagrangian whose bosonic sector is invariant under the global O(d,d) transformations J. E. Lidsey and J. Maharana [8] have defined the following superfields

$$m_{ij}(\tau,\theta,\bar{\theta}) \equiv M_{ij}(\tau) + i\bar{\psi}_{ij}(\tau)\theta + i\psi_{ij}(\tau)\bar{\theta} + F_{ij}(\tau)\theta\bar{\theta} \quad (2.14)$$

$$D(\tau,\theta,\bar{\theta}) \equiv \bar{\phi}(\tau) + i\bar{\chi}(\tau)\theta + i\chi(\tau)\bar{\theta} + f(\tau)\theta\bar{\theta} \quad (2.15)$$

where $M_{ij}(\tau)$ is given by (2.5), $\{\psi_{ij}, \chi\}$ are complex spinors and (i,j) = (1, 2, . . . , 2d). Furthermore, the N = 2 supersymmetric effective action [8] is written as

$$I_{Susy}^{N=2} \equiv \int d\tau \int d\theta\, d\bar{\theta}\left(\sum + Y\right) \quad (2.16)$$

where

$$\sum(\tau,\theta,\bar{\theta}) \equiv \frac{1}{8}\hat{D}_1 m_{ij}\eta^{jk}\hat{D}_2 m_{kl}\eta^{li} \quad (2.17)$$

$$Y(\tau,\theta,\bar{\theta}) \equiv \frac{1}{8}\hat{D}_1 D\,\hat{D}_2 D - W(D) \quad (2.18)$$

with $\hat{D}_1, \hat{D}_2$ the derivative operators namely

$$\hat{D}_1 \equiv -\frac{\partial}{\partial\bar{\theta}} + i\theta\frac{\partial}{\partial\tau} \quad (2.19)$$

$$\hat{D}_2 \equiv \frac{\partial}{\partial\theta} - i\bar{\theta}\frac{\partial}{\partial\tau} \quad (2.20)$$

and the potential W(D) is an arbitrary function of D. The expansion of the potential W(D) around $\bar{\phi}$ and the use of the expressions (2.14) and (2.15) in (2.17) and (2.18) leads respectively to

$$I_{Susy}^{N=2} = \int d\tau (L_g + L_1) \qquad (2.21)$$

where

$$L_g = \frac{1}{8}\left[i\psi_{ij}\eta^{jk}\overline{\psi}'_{kl}\eta^{li} - i\psi'_{ij}\eta^{jk}\overline{\psi}_{kl}\eta^{li} + iM'_{ij}\eta^{jk}M'_{kl}\eta^{li}\right] \qquad (2.22)$$

$$L_1 = (\bar{\phi}')^2 + \frac{i}{2}(\bar{\chi}\chi' - \bar{\chi}'\chi) + \frac{1}{2}f^2 - \frac{1}{\sqrt{2}}f(\partial_{\bar{\phi}}W) - \sqrt{2}(\partial_{\bar{\phi}}^2 W)\bar{\phi}f - \frac{1}{4}(\partial_{\phi}^2 W)[\bar{\chi},\chi]_- \qquad (2.23)$$

We remark that in the expression of $L_1$ given by J. E. Lidsey and J. Maharana [8] the coupling term

$$(\partial_{\bar{\phi}}^2 W)\bar{\phi} f \qquad (2.24)$$

had been omitted. In this case the use of the equation of motion for the auxiliary field f gives

$$f = \frac{1}{\sqrt{2}}\partial_{\bar{\phi}}W \qquad (2.25)$$

and the N = 2 supersymmetric action is as follows

$$I_{Susy}^{N=2} = \int d\tau \left\{ \frac{1}{8}\left(i\psi_{ij}\eta_{jk}\overline{\psi}'_{kl}\eta^{li} - i\psi'_{ij}\eta^{jk}\overline{\psi}_{kl}\eta^{li} + M'_{ij}\eta^{jk}M'_{kl}\eta^{li}\right) + (\bar{\phi}')^2 + \frac{i}{2}(\bar{\chi}\chi' - \bar{\chi}'\chi) - \frac{1}{4}(\partial_{\bar{\phi}}W)^2 - \frac{1}{4}(\partial_{\phi}^2 W)[\bar{\chi},\chi]_- \right\} \qquad (2.26)$$

From this action, the classical Hamiltonian is written as

$$H = 2\pi_{ij}\eta^{jk}\pi_{kl}\eta^{li} + \frac{1}{4}\pi_{\phi}^2 + \frac{1}{4}(\partial_{\bar{\phi}}W)^2 + \frac{1}{4}(\partial_{\phi}^2 W)[\bar{\chi},\chi]_- \qquad (2.27)$$

where

$$\pi_{ij} \equiv \pi_{M_{ij}} \qquad (2.28)$$

An identical expression for the Hamiltonian (2.27) is obtained from the anticommutator of the supercharges $Q$ and $\overline{Q}$, which are defined by

$$Q \equiv 2\pi_{ij}\eta^{jk}\psi_{kl}\eta^{li} + \frac{1}{\sqrt{2}}(\pi_{\bar{\phi}} + i\partial_{\bar{\phi}}W)\chi$$

$$\overline{Q} \equiv 2\pi_{mn}\eta^{nr}\overline{\psi}_{rp}\eta^{pm} + \frac{1}{\sqrt{2}}(\pi_{\bar{\phi}} - i\partial_{\bar{\phi}}W)\overline{\chi}$$

(2.29)

with

$$Q^2 = 0 = \overline{Q}^2,$$ (2.30)

namely

$$\{Q, \overline{Q}\} = 2H$$ (2.31)

$$[H, Q] = 0 = [H, \overline{Q}]$$ (2.32)

Therefore, there exists an N = 2 supersymmetric in quantum cosmology [13,7] which can be considered as a direct extension of the O(d,d) T-duality of the toroidally compactified string effective action (2.1). In the next section, we show that there exists even N = 4 supersymmetry in the quantum cosmology.

3 - N = 4 Supersymmetric String Quantum Cosmology

It is known that a hidden symmetry exists in all Bianchi class A models [14] where the classical superspace Hamiltonian may be viewed as the bosonic part of a supersymmetric Hamiltonian [7]. This implies that a supersymmetry can be introduced at the quantum level. These supersymmetric extensions of the quantum theory have significant consequences for quantum cosmology and may provide valuable insight into some of the questions relevant to a complete theory of quantum gravity [15]. However, in order to obtain an N = 4 supersymmetric lagrangian whose bosonic sector is invariant under the global O(d,d) transformations we consider the generators for the N = 4 supersymmetry which are defined by

$$\hat{Q}^+ \equiv \frac{\partial}{\partial \theta^-} - \theta^+ \partial_\tau$$
$$\hat{Q}^- \equiv \frac{\partial}{\partial \theta^+} - \theta^- \partial_\tau \tag{3.1}$$

where

$$\theta^\pm \equiv \theta^{\pm\beta} \equiv \theta^{\alpha\beta} U_\alpha^\pm,$$
$$U^{\pm\alpha} U_\alpha^\mp = 1, \quad U^{\pm\alpha} U_\alpha^\pm = 0 \tag{3.2}$$

with β and $U_\alpha^\pm, \alpha = 1, 2$ are the harmonic variables [16]. The N = 4 supersymmetric transformation of a superfield $\Omega$ is given by

$$\delta\Omega = \left(\varepsilon_1^+ \hat{Q}^- + \varepsilon_2^+ \hat{Q}^-\right)\Omega \tag{3.3}$$

where $\varepsilon_i^\pm$ are arbitrary superparameters. Furthermore, we define the following N = 4 superfields

$$m_{ij}(\tau, \theta^+, \theta^-, U) \equiv M_{ij}(\tau, U) + \theta^+ \psi_{ij}^-(\tau, U) + \theta^- \psi_{ij}^+(\tau, U)$$
$$+ \theta^+ \theta^- H_{ij}(\tau, U) + \theta^{+2} h_{ij}^{--}(\tau, U) + \theta^{-2} h_{ij}^{++}(\tau, U) \tag{3.4}$$
$$+ \theta^+ \theta^{-2} K_{ij}^+(\tau, U) + \theta^- \theta^{+2} K_{ij}^-(\tau, U) + \theta^{+2} \theta^{-2} F_{ij}(\tau, U)$$

$$D(\tau, \theta^+, \theta^-, U) \equiv \bar{\phi}(\tau, U) + \theta^+ \chi^-(\tau, U) + \theta^- \chi^+(\tau, U)$$
$$+ \theta^+ \theta^- d(\tau, U) + \theta^{+2} b^{--}\tau, U) + \theta^{-2} b^{++}(\tau, U) \tag{3.5}$$
$$+ \theta^+ \theta^{-2} C^+(\tau, U) + \theta^- \theta^{+2} C^-(\tau, U) + \theta^{+2} \theta^{-2} f(\tau, U)$$

in terms of which the N = 4 supersymmetric effective action is expressed

$$I_{Susy}^{(4,0)} = \int d\tau \int d^2\theta^+ d^2\theta^- dU \left[\sum + Y\right] \tag{3.6}$$

where $\Sigma$ and Y are respectively given by

$$\sum \equiv \frac{1}{8}(\hat{D}^{++} m_{ij})\eta^{jk}(\hat{D}^{--} m_{kl})\eta^{li} + 2g_1(\hat{D}^{-2} m_{ij})\eta^{jk}(\hat{D}^{+2} m_{kl})\eta^{li} \tag{3.7}$$

$$Y \equiv (\hat{D}^{++}D)(\hat{D}^{--}D) - \frac{1}{4}g_2(\hat{D}^{-2}D)(\hat{D}^{+2}D) - W(D) \tag{3.8}$$

where $g_1$ and $g_2$ are coupling constants, the potential W(D) is an arbitrary function of $D$ and the derivative operators are given by

$$\hat{D}^- = \frac{\partial}{\partial \theta^+} + \theta^- \partial_\tau$$
$$\hat{D}^+ = \frac{\partial}{\partial \theta^-} + \theta^+ \partial_\tau \tag{3.9}$$

$$\hat{D}^{++} = \partial^{++} + \theta^+ \theta^+ \partial_\tau$$
$$\hat{D}^{--} = \partial^{--} + \theta^- \theta^- \partial_\tau \tag{3.10}$$

where

$$\partial^{\pm\pm} = U^{\pm i} \frac{\partial}{\partial U^{\mp i}}$$

are the harmonic derivatives [16].

By using the expansions (3.4), (3.5) and by expanding the potential W(D) around $\bar{\phi}$ the (4,0) supersymmetric effective action becomes

$$I_{Susy}^{(4,0)} = \int d\tau dU \left\{ \frac{1}{8} \left[ M'_{ij} \eta^{jk} M'_{kl} \eta^{li} + \psi_{ij}^+ \eta^{jk} \psi_{kl}^{-'} \eta^{li} - \psi_{ij}^{+'} \eta^{jk} \psi_{kl}^- \eta^{li} \right. \right.$$
$$\left. + g_1 K_{ij}^{+'} \eta^{jk} K_{kl}^- \eta^{li} - g_1 K_{ij}^+ \eta^{jk} K_{kl}^{-'} \eta^{li} \right]$$
$$+ \bar{\phi}^{'2} + \frac{1}{2}(\chi^+ \chi^{-'} - \chi^{+'} \chi^-) + \frac{1}{2}g_2(C^+ C^{-'} - C^{+'} C^-) \tag{3.11}$$
$$\left. - \frac{\partial^2 W}{\partial \bar{\phi}^2}(\chi^+ C^- - C^+ \chi^-) - \frac{1}{4g_2}\left[ \frac{\partial W}{\partial \bar{\phi}} + \bar{\phi} \frac{\partial^2 W}{\partial \bar{\phi}^2} \right]^2 + ... \right\}$$

where we have used the equations of motion of auxiliary superfields $(h_{kl}^{\mp\mp}, F_{kl}, H_{kl})$ and $(d, b^{\pm\pm}, f)$ namely

$$2\partial_\tau \partial^{\pm\pm} M_{ij} + \frac{1}{2}(\partial^{++}\partial^{--} + \partial^{--}\partial^{++})h_{ij}^{\pm\pm} = 0$$

$$(\partial^{++}\partial^{--} + \partial^{--}\partial^{++}) M_{ij} = 0 \qquad (3.12)$$

$$\partial^{++}\partial^{--} H_{ij} = 0$$

$$\left(\partial^{++}\partial^{--} + \frac{\partial^2 W}{\partial \bar{\phi}^2}\right)d = 0$$

$$\left(\partial^{\pm\pm}\partial^{\mp\mp} + \frac{1}{4}\frac{\partial^2 W}{\partial \bar{\phi}^2}\right)b^{\pm\pm} = -2\partial^{\pm\pm}\partial_\tau \bar{\phi} \qquad (3.13)$$

$$f = -\frac{1}{2g_2}\left(\frac{\partial W}{\partial \bar{\phi}} + \frac{\partial^2 W}{\partial \bar{\phi}^2}\right)\bar{\phi}$$

and where the dotted line in (3.11) indicates couplings between the superfields $(\psi_{ij}^\pm, K_{ij}^\mp), (\chi^\pm, C^\mp)$ and higher derivative terms with respect to $\tau$ and $\bar{\phi}$ have been omitted. Let us note that we have taken into account the coupling $\bar{\phi}\,\partial^2_{\bar{\phi}} W$ that had been ignored in the N = 2 supersymmetric case [8]. Furthermore, concerning the integration with respect to the harmonic variables we consider the following harmonic expansion of each component of the superfields (3.4) and (3.5) in terms of symmetrised products of $U^\pm$ [16]

$$g^{(q)}(\tau, U) = \sum_{n=0}^{\infty} g^{(\alpha_1 \ldots \alpha_{n+q} \beta_1 \ldots \beta_n)}(\tau)\, U^+_{(\alpha_1} \ldots U^+_{\alpha_{n+q}} U^-_{\beta_1} \ldots U^-_{\beta_n)} \qquad (3.14)$$

where q is the Cartan-Weyl charge. The integration over $U^\pm$ is defined by using the rules

$$\int dU\, 1 = 1,$$
$$\int dU\, U^+_{(\alpha_1} \ldots U^+_{\alpha_n} U^-_{\beta_1} \ldots U^-_{\beta_m)} = 0,\; n+m > 0 \qquad (3.15)$$

On the other hand, the action (3.11) reduces to the bosonic action (2.2) in the limit where the Grassmann variables vanish and if we identify the potential

$$\left(\frac{\partial W}{\partial \bar{\phi}} + \bar{\phi}\frac{\partial^2 W}{\partial \bar{\phi}^2}\right)^2 = -4g_2 V e^{-2\bar{\phi}} \qquad (3.16)$$

In the remaining part of this section we take for simplicity $g_1=1, g_2=1$ and in order to derive the classical Hamiltonian for the N = 4 supersymmetric case let us give the classical momenta conjugate to the bosonic and fermionic degrees of freedom in action (3.11) namely

$$\pi_{M_{mn}} = \frac{\partial L}{\partial M'_{mn}} = \frac{1}{4}\eta^{nk} M'_{kl} \eta^{lm}$$

$$\pi_{\psi^+_{mn}}^- = \frac{\partial L}{\partial \psi^{+'}_{mn}} = -\frac{1}{8}\eta^{nk} \psi^-_{kl} \eta^{lm}$$

$$\pi_{\psi^-_{mn}}^+ = \frac{\partial L}{\partial \psi^{-'}_{mn}} = -\frac{1}{8}\psi^+_{ij} \eta^{jm} \eta^{ni}$$

$$\pi_{K^+_{mn}}^- = \frac{\partial L}{\partial k^{+'}_{mn}} = \frac{1}{8}\eta^{nk} k^-_{kl} \eta^{lm}$$

$$\pi_{K^-_{mn}}^+ = \frac{\partial L}{\partial k^{-'}_{mn}} = \frac{1}{8}k^+_{ij} \eta^{jm} \eta^{ni}$$

$$\pi_{\bar{\phi}} = \frac{\partial L}{\partial \bar{\phi}'} = 2\bar{\phi}' \qquad (3.17)$$

$$\pi_{\chi^+}^- = \frac{\partial L}{\partial \chi^{+'}} = -\frac{1}{2}\chi^-$$

$$\pi_{\chi^-}^+ = \frac{\partial L}{\partial \chi^{-'}} = -\frac{1}{2}\chi^+$$

$$\pi_{C^+}^- = \frac{\partial L}{\partial C^{+'}} = -\frac{1}{2}C^-$$

$$\pi_{C^-}^+ = \frac{\partial L}{\partial C^{-'}} = -\frac{1}{2}C^+$$

with

$$I^{(4,0)}_{Susy} = \int d\tau\, dU\, L$$

Therefore, the N = 4 supersymmetric classical Hamiltonian is as follows

$$H = M'_{ij}\pi_{M_{ij}} + \psi^{+'}_{ij}\pi^-_{\psi^+_{ij}} + \psi^{-'}_{ij}\pi^+_{\psi^-_{ij}} + K^{+'}_{ij}\pi^-_{k^+_{ij}} + K^{-'}_{ij}\pi^+_{k^-_{ij}} + \bar{\phi}'\pi_{\bar{\phi}} + \chi^{+'}\pi^-_{\chi^+} + \chi^{-'}\pi^+_{\chi^-} + C^{+'}\pi^-_{C^+} + C^{-'}\pi^+_{C^-} - L \qquad (3.18)$$

which can be rewritten by using (3.17) as

$$H = 2\pi_{ij}\eta^{jk}\pi_{kl}\eta^{li} + \frac{1}{4}(\pi_{\bar{\phi}})^2 + \frac{\partial^2 W}{\partial \bar{\phi}^2}(\chi^+ C^- - C^+ \chi^-)$$
$$+ \frac{1}{4}\left[\frac{\partial W}{\partial \bar{\phi}} + \bar{\phi}\frac{\partial^2 W}{\partial \bar{\phi}^2}\right]^2$$
(3.19)

Then, the bosonic component of the N = 4 supersymmetric Hamiltonian (3.19) corresponds to the classical Hamiltonian for the cosmological model (2.10). The fermions do not appear in the expression of the Hamiltonian for the matrix $M_{ij}$ as in the N =2 supersymmetric case [8]. However, the N = 4 supersymmetric model is quantified by using the standard operator realization (2.12). Furthermore, we impose the following spinor algebra

$$\begin{aligned}
&\{\chi^\pm, \chi^\pm\} = 0, \qquad \{\chi^+, \chi^-\} = 1 \\
&\{C^\pm, C^\pm\} = 0, \qquad \{C^+, C^-\} = 1 \\
&\{\psi_{ij}^\pm, \psi_{kl}^\pm\} = 0, \qquad \{\psi_{ij}^+, \psi_{kl}^-\} = \eta_{ik}\eta_{jl} \\
&\{K_{ij}^\pm, K_{kl}^\pm\} = 0, \qquad \{K_{ij}^+, K_{kl}^-\} = \eta_{ik}\eta_{jl} \\
&\{\chi_{ij}^\pm, \psi_{kl}^\pm\} = 0 = \{\chi_{ij}^\pm, K_{kl}^\pm\}, \dots
\end{aligned}$$
(3.20)

which is satisfied by introducing the set of the Grassmann variables $\{\varsigma_{ij}^\pm, \xi_{ij}^\pm, \beta^\pm, \mu^\pm\}$ such that

$$\psi_{kl}^+ = \eta_{kp}\frac{\partial}{\partial \varsigma_{pr}^-}\eta_{rl} \quad , \quad \psi_{ij}^- = \varsigma_{ij}^-$$

$$K_{kl}^+ = \eta_{kp}\frac{\partial}{\partial \xi_{pr}^-}\eta_{rl} \quad , \quad K_{ij}^- = \xi_{ij}^-$$

$$\chi^+ = \frac{\partial}{\partial \beta^-} \quad , \quad \chi^- = \beta^-$$

$$C^+ = \frac{\partial}{\partial \mu^-} \quad , \quad C^- = \mu^-$$
(3.21)

An identical expression for the Hamiltonian (3.19) can be obtained from the anticommutator of the supercharges which are defined by

$$Q^+ = \pi_{ij}\eta^{jk}\left[\psi^+_{kl} + K^+_{kl}\right]\eta^{li} + \frac{1}{\sqrt{2}}\left[\pi_{\bar\phi}\chi^+ + i\left(\partial_{\bar\phi}W + \bar\phi\frac{\partial^2 W}{\partial\bar\phi^2}\right)C^+\right]$$

$$Q^- = \pi_{mn}\eta^{nr}\left[\psi^-_{rp} + K^-_{rp}\right]\eta^{pm} + \frac{1}{\sqrt{2}}\left[\pi_{\bar\phi}\chi^- - i\left(\partial_{\bar\phi}W + \bar\phi\frac{\partial^2 W}{\partial\bar\phi^2}\right)C^-\right]$$

(3.22)

**These supercharges satisfy the following relations**

$$Q^{+2} = 0 = Q^{-2} \tag{3.23}$$

**and lead to**

$$\{Q^+, Q^-\} = 2H$$
$$[H, Q^+] = 0 = [H, Q^-] \tag{3.24}$$

**Thus, there exists an N = 4 supersymmetry in the quantum cosmology as in the N = 2 supersymmetric case [7,13]. Defining the conserved fermion number solves the (4,0) supersymmetric quantum constraints**

$$F = \psi^-_{ij}\eta^{jk}\psi^+_{kl}\eta^{li} + K^-_{ij}\eta^{jk}K^+_{kl}\eta^{li} + \chi^-\chi^+ + C^-C^+ \tag{3.25}$$

**which satisfies**

$$[Q^+, F] = 2Q^+$$
$$[Q^-, F] = -2Q^-$$
$$[H, F] = 0 \tag{3.26}$$

**Furthermore, the fermion vacuum $|0\rangle$ is defined as follows**

$$\psi^+_{ij}|0\rangle = 0 = K^+_{ij}|0\rangle, \qquad \forall i,j$$
$$\chi^+|0\rangle = 0 = C^+|0\rangle \tag{3.27}$$

**and the state with zero fermion number $|\psi_0\rangle$ is given by**

$$|\psi_0\rangle \equiv h(M_{ij}, \bar\phi)|0\rangle \tag{3.28}$$

where h is an arbitrary function of $M_{ij}$ and $\bar{\phi}$. Since the supersymmetry implies that the wave function of the Universe is annihilated by the supercharges $Q^{\pm}$ then the state (3.28) is automatically annihilated by the supercharge $Q^{+}$ and it is annihilated by the supercharge $Q^{-}$ if the following conditions are satisfied

$$\frac{\partial h}{\partial M_{ij}} = 0 \tag{3.29.a}$$

$$\left[\frac{\partial h}{\partial \bar{\phi}}\chi^{-} + \left(\frac{\partial W}{\partial \bar{\phi}} + \bar{\phi}\frac{\partial^2 W}{\partial \bar{\phi}^2}\right)h C^{-}\right]|0\rangle = 0 \tag{3.29.b}$$

The condition (3.29) shows that

$$h = h(\bar{\phi}) \tag{3.30}$$

and if we assume that

$$\chi^{-}|0\rangle \neq C^{-}|0\rangle \tag{3.31}$$

the condition (3.29.b) leads to

$$\frac{\partial h}{\partial \bar{\phi}} = 0 \tag{3.32}$$

which means that h = constant and

$$\left(\frac{\partial W}{\partial \bar{\phi}} + \bar{\phi}\frac{\partial^2 W}{\partial \bar{\phi}^2}\right) = \frac{\partial}{\partial \bar{\phi}}\left[\bar{\phi}\frac{\partial W}{\partial \bar{\phi}}\right] = 0 \tag{3.33}$$

which implies, with the use of (3.16), that this case corresponds to a situation where the potential V = 0. Moreover, if we take

$$\chi^{-}|0\rangle = C^{-}|0\rangle \tag{3.34}$$

then, condition (3.29.b) becomes

$$\frac{\partial h}{\partial \bar{\phi}} + \frac{\partial}{\partial \bar{\phi}}\left[\bar{\phi}\frac{\partial W}{\partial \bar{\phi}}\right]h = 0 \tag{3.35}$$

where the general solution is given by

$$|\psi_0\rangle = e^{-W_1(\bar{\phi})}|0\rangle \tag{3.36}$$

with

$$W_1(\bar{\phi}) = \bar{\phi}\frac{\partial W}{\partial \bar{\phi}} \tag{3.37}$$

This solution is uniquely determined by the potential (3.16) as in the N = 2 supersymmetric case [8].

On the other hand, we define the one-fermion state $|\psi_1^-\rangle$ as follows

$$|\psi_1^-\rangle = \left[h_{ij}\eta^{jk}\psi_{kl}^-\eta^{li} + k_{ij}\eta^{jk}K_{kl}^-\eta^{li} + h_\chi \chi^- + k_C C^-\right]e^{-W_1(\bar{\phi})} = 0 \tag{3.38}$$

where $h_{ij}$, $k_{ij}$, $h_\chi$, $k_C$ are arbitrary functions of the bosonic variables over the configuration space. Such state is annihilated by the supercharge $Q^-$ if we consider

$$\begin{aligned} h_{ij} &= k_{ij} = \pi_{ij} h \\ h_\chi &= \frac{1}{\sqrt{2}}\pi_{\bar{\phi}} h \\ k_C &= -\frac{i}{\sqrt{2}}\left(\frac{\partial W}{\partial \bar{\phi}} + \bar{\phi}\frac{\partial^2 W}{\partial \bar{\phi}^2}\right)h \end{aligned} \tag{3.39}$$

Therefore, the state $|\psi_1^-\rangle$ is rewritten as

$$|\psi_1^-\rangle = Q^- h\, e^{-W_1(\bar{\phi})}|0\rangle \tag{3.40}$$

and the action of the supercharge $Q^+$ on it, with the use of equation (3.24), leads to

$$Q^+|\psi_1^-\rangle = 2H\, h\, e^{-W_1(\bar{\phi})}|0\rangle \tag{3.41}$$

This equation is satisfied if the function $h(M_{ij}, \bar{\phi})$ is a solution of the following equation

$$\left[2\frac{\delta}{\delta M_{ij}}\eta^{jk}\frac{\delta}{\delta M_{kl}}\eta^{li} + \frac{1}{4}\frac{\partial^2}{\partial\bar{\phi}^2} - \frac{1}{2}\left(\frac{\partial W_1}{\partial\bar{\phi}}\right)\frac{\partial}{\partial\bar{\phi}} - \frac{\partial^2 W_1}{\partial\bar{\phi}^2}\right]h(M_{ij},\bar{\phi}) = 0 \quad (3.42)$$

**Finally, for separable solution**

$$h(M_{ij}, \bar{\phi}) \equiv X(M_{ij}) Y(\bar{\phi}) \quad (3.43)$$

**the differential equation (3.42) implies that**

$$\left[\frac{\delta}{\delta M_{ij}}\eta^{jk}\frac{\delta}{\delta M_{kl}}\eta^{li} + \frac{C^2}{8}\right]X(M_{ij}) = 0 \quad (3.44)$$

$$\left[\frac{\partial^2}{\partial\bar{\phi}^2} - 2\left(\frac{\partial W_1}{\partial\bar{\phi}}\right)\frac{\partial}{\partial\bar{\phi}} - 4\frac{\partial^2 W_1}{\partial\bar{\phi}^2} - C^2\right]Y(\bar{\phi}) = 0 \quad (3.45)$$

where C is a separation constant. We note that for a constant dilaton potential equation (3.45) can be solved. However, the duality symmetries discussed in ref. [17] can be used for the N = 4 supersymmetric quantum cosmology model. Moreover, it is known that spatially flat isotropic cosmologies derived from the Brans-Dicke gravity action exhibit a scale factor duality invariance [18,6]. This classical duality, associated with the hidden N = 2 supersymmetry at the quantum level [18,6], can also be studied for the N = 4 supersymmetric case. Finally, the dualities of the Bianchi models related to N = 4 supersymmetry would be of interest to be investigated.

## 4 - Conclusion

In this paper, we have derived an N = 4 supersymmetric quantum cosmology from the N= 4 supersymmetric effective action. The N = 4 supersymmetric Hamiltonian operator is obtained by using the momenta conjugate to the bosonic and fermionic degrees of freedom. In the classical limit this operator reduces to the O(d,d) invariant Hamiltonian. The constraints on the wave function of the Universe are imposed by supersymmetry and imply that it should be annihilated by supercharges. On the other hand, the solutions of the N = 4 supersymmetric constraints are found for the zero and one-fermion states. We have seen that when the fermionic states $\chi^-|0\rangle$ and $C^-|0\rangle$ are not identical the potential $V(\bar{\phi})$ has to be cancelled. Moreover, for the case where these two states are equivalent the Wheeler-DeWitt equation is obtained and differs from the N = 2 supersymmetric case just by the second derivative of the potential $W_1(\bar{\phi})$ which has been taken into account.

## Acknowledgments

The authors would like to thank Professor K. S. Narain for the interesting discussions and for reading the manuscript and Professor M. Virasoro, the International Atomic Energy Agency and UNESCO for hospitality at the Abdus Salam International Centre for Theoretical Physics, Trieste. This work is supported by the program of the Associate and Federation Schemes of the Abdus Salam International Centre for Theoretical Physics, Trieste, Italy.

Quantum. Grav. **11**, 1961 (1994).

[15] O. Obregon, J. Socorro and J. Benitez, Phys. Rev. **D47**, 4471 (1993); J. Bene and R. Graham, **D49**, 799 (1994).

[16] A. Galperin, E. Ivanov, S. Kalitzin, V. Ogievetsky and E. Sokatchev; Class. Quantum. Grav. **1**, 469 (1984).

[17] J. Maharana, Phys. Lett. **B296**, 65 (1992).

[18] Dominic Clancy, J. E. Lidsey and Reza Tavakol, Class. Quantum. Grav. **15**, 257 (1998).